\documentclass[prl,twocolumn,showpacs,amsmath,amssymb,preprintnumbers]{revtex4}

\usepackage{graphicx}
\usepackage{multirow}
\usepackage{amsmath}
\usepackage{feyn}

\font\FermiSmallfont=cmssq8 scaled 1200

\def\LANLppthead#1{
\null 
\begin{center}\vskip -1.0truein{\hbox to 7.5truein {
\hfill
\vbox to 1in {\vfill \FermiSmallfont
              \hbox{#1}
              \vfill}
}}\vskip-0.0truein\end{center}}

\begin{document}

\title{Prospects for Neutrino-Antineutrino Transformation in Astrophysical Environments}
\preprint{LA-UR-14-24562}

\author{Alexey Vlasenko$^{1,3}$}
\author{George M. Fuller$^{1,3}$}
\author{Vincenzo Cirigliano$^{2,3}$}

\affiliation{$^{1}$Department of Physics, University of California, San Diego, La
Jolla, California 92093, USA}
\affiliation{$^{2}$Theoretical Division, Los Alamos National Laboratory, Los Alamos,
New Mexico 87545, USA}
\affiliation{$^{3}$Neutrino Engineering Institute, New Mexico Consortium, Los
Alamos, New Mexico 87545, USA}

\date{June 21, 2014}





\begin{abstract}



We examine whether the newly derived neutrino spin coherence could lead to large-scale coherent neutrino-antineutrino conversion. In a linear analysis we find that such transformation is largely suppressed, but demonstrate that nonlinear feedback can enhance it.  We point out that conditions which favor this feedback may exist in core collapse supernovae and in binary neutron star mergers.

\end{abstract}

\pacs{14.60.Pq, 97.60.Bw, 26.50.+x, 26.30.+k, 13.15.+g}

\maketitle

In this letter we address the prospects for spin coherence, a recently revealed \cite{Vlasenko:2014lr,Cirigliano:2014lr}
aspect of medium-affected neutrino physics, to facilitate the inter-conversion of
neutrinos and antineutrinos in the core collapse supernova and compact object merger environments. 
This is important because the asymmetry between $\nu_e$ and
$\bar\nu_e$ fluxes and energy spectra in these sites can influence 
both dynamics and the neutron-to-proton ratio \cite{Qian93}, a key determinant of nucleosynthesis. 
The stakes are high because these are our best candidate sites for the origin of the heaviest elements.
Moreover, future neutrino \cite{de-Gouvea:2013yq,Gilchriese:2014vn} and gravitational radiation \cite{Punturo:2010rt} observations may give insights into these venues.  

Neutrino spin coherence was discovered in the course of deriving the quantum kinetic
equations (QKEs) that govern the evolution of neutrino distributions in a medium of
matter and neutrinos \cite{Vlasenko:2014lr,Cirigliano:2014lr}. In that work, nonzero neutrino mass and the presence of anisotropy in the matter or in the neutrino fields were shown to be necessary for coherent transformation between left-handed and right-handed neutrino states.  In short, the QKEs show that in an anisotropic medium the neutrino propagation states (energy states) can be coherent mixtures of left- and right-handed ({\it i.e.,} neutrino and antineutrino) states. 

The QKEs for flavored particles have been considered in many contexts \cite{Raffelt:1993fj,Sigl:1993fr,Raffelt:1993kx,McKellar:1994uq,Sawyer:2005yg,Strack:2005fk,Cardall:2008lr,Herranen:2008qy,Herranen:2009fk,Gava:2009yq,Volpe:2013lr,Enqvist:1991yq,Barbieri:1991fj,Dodelson:1994rt,Shi:1996vn,Foot:1997rt,Bell:1999fk,Dolgov:2000fr,Volkas:2000uq,Abazajian:2001lr,Dolgov:2002ve,Kusenko:2005qy,Boyanovsky:2007fk,Boyanovsky:2007kx,Boyanovsky:2007lr,Kishimoto:2008pd,Kusenko:2009lr,Cirigliano:2010lr,Cirigliano:2011lr,Vlasenko:2014lr,Cirigliano:2014lr,Bhupal-Dev:2014fk}.  For neutrinos, the QKEs describe coherent forward evolution as well as scattering and thermalization, and can reduce to Schr\"odinger-like equations for flavor evolution or the Boltzmann equation in certain limits.
Compared to the standard Schr\"odinger-like treatment of neutrino flavor
transformation, as described, for example, in Ref.s~\cite{Wolfenstein78,Mikheyev85,Duan:2010fr,Fuller87,Notzold:1988fv,Sawyer:1990lr,Qian93,Samuel:1993sf,Qian95,Samuel:1996rm,Pastor:2002zl,Pastor02,Balantekin05,Fuller06,Duan06a,Duan06b,Duan06c,Hannestad:2006qd,Duan07a,Duan07b,Balantekin:2007kx,Duan07c,Duan08,Kneller:2008rt,Lunardini08,Dasgupta:2008kx,Gava:2009yq,Dasgupta:2011uq,Friedland:2010yq,Duan:2011lr,Mirizzi:2012qy}, the QKEs contain two novel features.  One is 
the collision term, which allows exchange of particle number and flavor information
between neutrinos of different energies traveling on different trajectories.  The
other feature is helicity mixing, or spin coherence which,
for Majorana neutrinos, mediates exchange of information between neutrino and
antineutrino states.

In Ref.s.~\cite{Cherry:2012lr,Cherry:2013lr}, it was pointed out that even a small amount of non-forward neutrino scattering can potentially have large effects on supernova neutrino flavor transformation.  However, here we temporarily set aside the issue of collisions and retain only coherent forward scattering terms in the QKEs, and consider the question of whether the spin coherence terms can possibly be important in compact object environments.  In the absence of the collision term, the QKEs take the following form:
\begin{eqnarray}
D{\cal F}+i\left[{\cal H},{\cal F}\right] = 0
\end{eqnarray}
Here ${\cal F}$ is a density matrix for neutrino and antineutrino states, $D$ is a Vlasov derivative operator, and ${\cal H}$ is a Hamiltonian for the evolution of the density matrix.  For 3 flavors of neutrinos and anti-neutrinos, the density matrix is a $6\times 6$ Hermitean matrix with the structure
\begin{eqnarray}
{\cal F} = \left(\begin{array}{cc}f & \phi \\ \phi^\dagger & \bar{f}^T\end{array}\right)
\end{eqnarray}
where $f$ and $\bar{f}$ are the usual $3\times 3$ density matrices for neutrinos and antineutrinos and $\phi$ encodes coherence between neutrinos and antineutrinos.

In dense environments, where coherent forward scattering of neutrinos on the matter background and on other neutrinos gives the dominant contribution to the neutrino potential energy, the Hamiltonian consists of a leading-order contribution and a correction:
\begin{eqnarray}
{\cal H} = \left(\begin{array}{cc}H^{\left(1\right)} & 0 \\ 0 & -H^{\left(1\right)T}\end{array}\right)+\left(\begin{array}{cc}H^{\left(2\right)} & H_{\nu\bar{\nu}}^{\left(2\right)} \\ H_{\nu\bar{\nu}}^{\left(2\right)\dagger} & H^{\left(2\right)T}\end{array}\right).
\end{eqnarray}

The form of the QKEs in Eqn. (1) is similar to equations for the evolution of neutrinos with a magnetic moment in a strong magnetic field \cite{Dvornikov:2012lr,de-Gouvea:2012fk,de-Gouvea:2013lr}, but in the case of the QKEs, $\nu\rightleftharpoons\bar{\nu}$ mixing can occur without a magnetic field or a neutrino magnetic moment.

$H^{\left(1\right)}$ is $O\left(G_F\right)$ while terms that mediate $\nu\rightleftharpoons\bar{\nu}$ mixing, $H_{\nu\bar{\nu}}^{\left(2\right)}$, are $O\left(mG_F/E\right)$, , where $m$ is the neutrino mass and E is the neutrino energy.  For neutrinos in a supernova envelope, with reasonable assumptions about the neutrino mass and energy, $m/E\sim 10^{-7}-10^{-8}$.  Thus under generic conditions $H^{\left(2\right)}_{\nu\bar{\nu}}$ is negligible and we do not expect to see any significant helicity transformation.  However, in special conditions, the potential for a neutrino state can be close to that for an antineutrino state.  In this case, the behavior of the system can be dominated by the $\nu\rightleftharpoons\bar{\nu}$ mixing term.

The solution of the full QKEs is a difficult numerical problem.  To get an idea of the effects of helicity mixing, we construct a highly simplified toy model.  Spin coherence can occur even with one flavor, so to construct the simplest possible model we retain only the electron flavor neutrinos and antineutrinos.  We also omit second-order corrections to the derivative $D$ and to the Hamiltonian, with the exception of the helicity mixing term $H_{\nu\bar{\nu}}^{\left(2\right)}$.

Writing ${\cal F} = f_i\sigma_i + f_0 {\bf I}$ and ${\cal H} = H_i\sigma_i + H_0{\bf I}$, the simplified QKEs for the one-flavor model are:
\begin{eqnarray}
D f_3 - 2H_1 f_2 = 0
\nonumber\\
D f_1 + 2H_3 f_2 = 0
\nonumber\\
D f_2 + 2H_1 f_3 - 2H_3 f_1 = 0
\end{eqnarray}
Here we have defined the coordinate system in such a way that $H_2 = 0$, which is possible for any specific neutrino trajectory.  In the special case of an axially symmetric geometry, $H_2$ can be set to zero for all trajectories.   $D$ is the derivative along the neutrino world line.  Here we simply consider evolution along a single world line and take $D = \partial_s$, where $s$ is the distance traveled by the neutrino.  In the absence of the collision term, $D f_0 = 0$. 

For the purpose of the toy model, we consider an axially symmetric cone of neutrinos propagating at a fixed angle $u = \cos\theta$ with respect to the axis of symmetry.  This geometry is illustrated in Fig.~1.  The neutrinos undergo coherent forward scattering among themselves and with the matter background (electrons and nucleons).  The properties of the matter background are varied slowly in order to determine what conditions can lead to significant $\nu\rightleftharpoons\bar{\nu}$ transformation.

\begin{figure}
\includegraphics[width=0.8in]{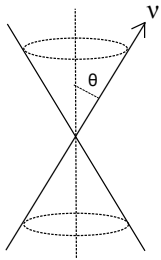}
\caption{Geometry of the single-flavor toy model.}
\label{drawing}
\end{figure}  

This problem corresponds to the behavior of crossed neutrino beams of infinite width in a time-varying background, and is somewhat different from that of neutrinos emitted from a spherically symmetric neutrino sphere.  In the latter case the angle $u$ changes along the neutrino path and the neutrinos are geometrically diluted as $1/r^2$ as they move outward.  These features of spherical geometry are straightforward to implement, but can complicate the problem and obscure the simple physical behavior that we wish to illustrate.

In the notation of Eq. (4), $f_3 = \left(f-\bar{f}\right)/2$, and, to leading order, $H_3 = H^{\left(1\right)}$. With this, $H_3$ is the Hamiltonian arising from coherent forward scattering of neutrinos with electrons, nucleons and other neutrinos.  Including the contribution from nucleons, the Hamiltonian for electron neutrinos is \cite{McLaughlin:1999fk, Abazajian:2001lr, Notzold:1988fv, Savage:1991qy}
\begin{eqnarray}
H_3 = \frac{G_F}{\sqrt{2}}\left(\left(3Y_e - 1\right)n_B+4\left(n_\nu-n_{\bar{\nu}}\right)-4 u J^r\right).
\end{eqnarray}
Here, $n_B$ is the baryon number density, $Y_e = {n_e}/{n_B}$ is the electron fraction, $J^r$ is the lepton number current along the axis of symmetry, and $n_\nu-n_{\bar{\nu}}$ is the lepton number density in neutrinos, given by
\begin{eqnarray}
n_\nu - n_{\bar{\nu}} =\int\frac{E'^2dE'}{2\pi^2}2f_3\left(E'\right).
\end{eqnarray}
The contribution to $J^r$ from neutrinos is $u\left(n_\nu-n_{\bar{\nu}}\right)$.  There can be additional contributions to $J^r$ from bulk motion of matter, such as infall or outflow.

$H_1$ is the $\nu\rightleftharpoons\bar{\nu}$ mixing part of the Hamiltonian, corresponding to $H_{\nu\bar{\nu}}^{\left(2\right)}$ in Eq. 3, and is equal to
\begin{eqnarray}
H_1 = 2\sqrt{2}G_F \sqrt{1-u^2}\frac{m J^r}{E}.
\end{eqnarray}
Neutrino-antineutrino mixing is large at {\it resonance}, {\it i.e.}, where $H_3\approx 0$, which occurs for
\begin{eqnarray}
 Y_e +\frac{4}{3}\left(Y_\nu-\frac{uJ^r}{n_B}\right) = \frac{1}{3}
\end{eqnarray}  
where $Y_\nu = \left(n_\nu-n_{\bar{\nu}}\right)/n_B$.  If the neutrino contribution to the Hamiltonian is relatively small, this corresponds to $Y_e\approx 1/3$, which can occur in or near the proto-neutron star (PNS) in a core collapse supernova \cite{Fischer:2010yq, Fischer:2012uq, Pons:1999fj, Liebendorfer:2003kx}, or near the central region of a compact object merger \cite{Perego:2014ys,Wanajo:2014fr}.  This is similar to the condition for active-sterile transformation in models with sterile neutrinos \cite{McLaughlin:1999fk,Wu:2014vn, Nunokawa:1997rt,Fetter:2003ys,Tamborra:2012fr,Balantekin:2003qy}.  

The single-flavor model is mathematically similar to the description of the Mikheyev-Smirnov-Wolfenstein (MSW) effect \cite{Wolfenstein78,Mikheyev85}.  Suppose that, in Eq. (5), $Y_e=Y_{e0}$ gives $H_3=0$.  We then begin with $Y_e < Y_{e0}$ and increase it to $Y_e>Y_{e0}$.  This situation can occur in a supernova when neutrinos pass from regions of low $Y_e$ inside the proto-neutron star to regions of higher $Y_e$ in the envelope.  For $Y_e<Y_{e0}$, $H_3$ is negative and neutrinos have a lower potential energy than antineutrinos.  For $Y_e>Y_{e0}$, $H_3$ is positive and the antineutrinos have lower potential energy.  This is a level crossing, schematically illustrated in Fig.~2, with instantaneous energy eigenvalues $E_\pm$.  A level crossing can also be achieved by varying $n_B$, $u$ and the neutrino distributions, and in a supernova all these quantities vary with radius.

If additional neutrino flavors are present, there are additional level crossings.  The matter potential for muon and tau neutrinos is $\left(G_F/\sqrt{2}\right)\left(Y_e-1\right)n_B$ \cite{McLaughlin:1999fk, Abazajian:2001lr, Notzold:1988fv, Savage:1991qy}, which leads to a level crossing between $\nu_e$ and $\bar{\nu}_{\mu, \tau}$ (and $\bar{\nu}_e$ and $\nu_{\mu,\tau}$) near $Y_e = 1/2$.  Similarly, a cancelation between the matter and the neutrino potentials could lead to a level crossing between $\nu_{\mu,\tau}$ and $\bar{\nu}_{\mu,\tau}$.  These level crossings are more likely to occur in the supernova envelope, where $Y_e$ and $Y_\nu$ can be relatively high \cite{Mezzacappa:2005lr,Fischer:2010yq}.

\begin{figure}
\includegraphics[width=2.8in]{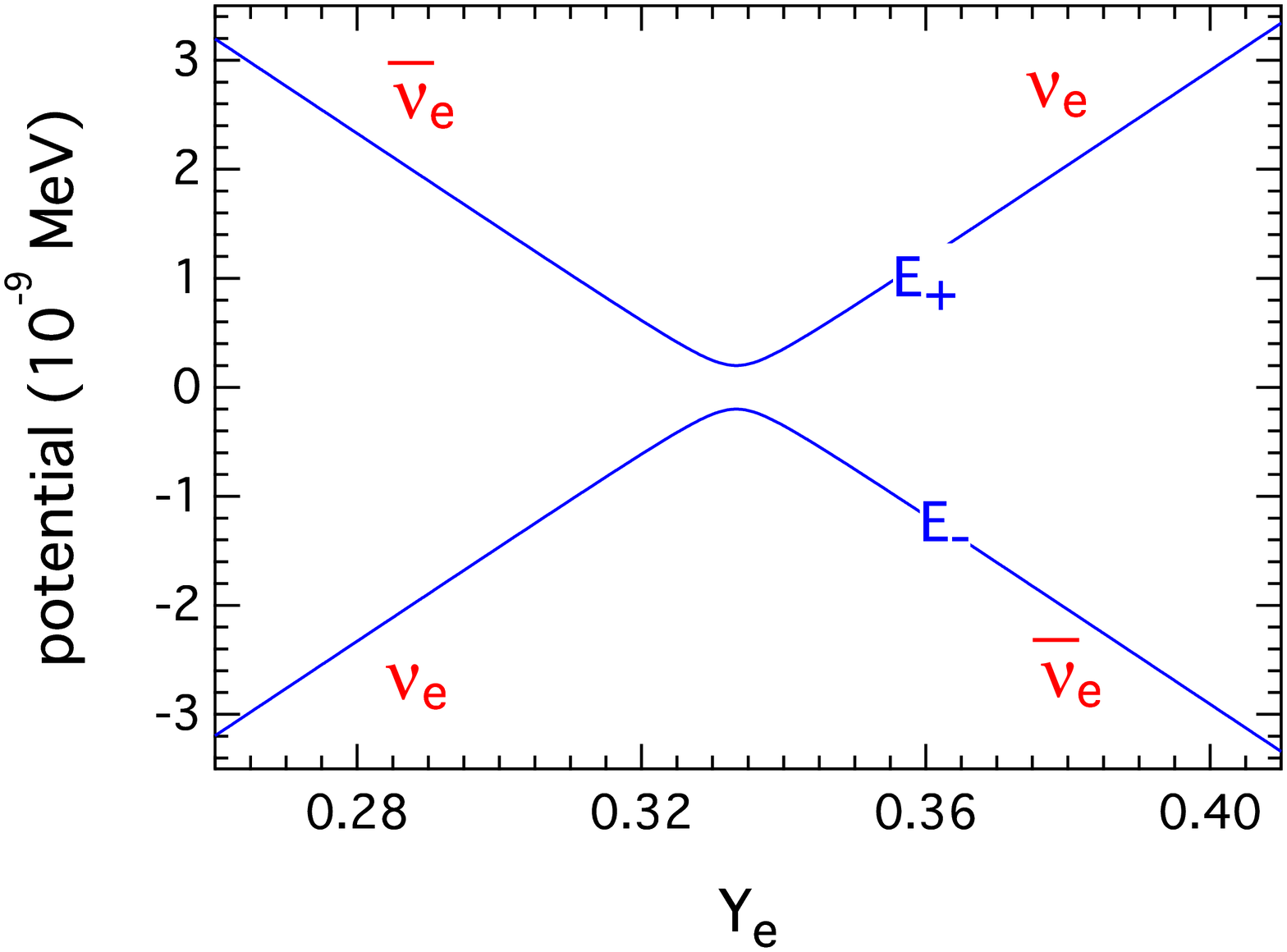}
\caption{Schematic level crossing diagram for $\nu_e\leftrightharpoons\bar{\nu}_e$ transformation. The potential $E_\pm = \pm\sqrt{H_3^2+H_1^2}$ is plotted against electron fraction $Y_e$, with off-diagonal potential $H_1$ exaggerated to clearly show the gap, $\left(E_+-E_-\right)|_{res}=2\left|H_1\right|$, at resonance.  Here neutrino contributions to $H_3$ are neglected.}
\label{lvcr}
\end{figure}

Provided that $H_3$ varies slowly enough (adiabatically), a neutrino that begins in the lower-energy state will remain in the lower-energy state, and therefore transform into an antineutrino.  Whether this transformation occurs is governed by the adiabaticity parameter, $\gamma = 2 H_1^2/\dot{H}_3$.  Because $H_1$ is generically smaller than $H_3$ by a factor of $m/E\approx 10^{-7}-10^{-8}$, the adiabaticity parameter for $\nu\rightleftharpoons\bar{\nu}$ transformation, which is proportional to $\left(m/E\right)^2$, is typically very small.  Unless $H_3$ varies extremely slowly, only neutrinos at very low energies can transform.  

However, the above analysis neglects effects of nonlinearity due to the dependence of the Hamiltonian on neutrino distributions.  To determine these effects, we discretize the energy and numerically solve our toy model using initial neutrino occupation numbers given by $1/\left(1+e^{\frac{E-\mu}{T}}\right)$, with $T = 4$ MeV and $\mu = 8$ MeV.  For simplicity, we start with a pure neutrino spectrum, hold the angle fixed at $u = 1/\sqrt{2}$ and the baryon number density at $n_B = 300\ {\rm MeV}^3$.  These values are roughly consistent with conditions above the neutrino sphere in a supernova \cite{Fischer:2010yq, Fischer:2012uq}.  We vary $Y_e$ as a function of distance $s$ traveled by neutrinos, as follows:
\begin{eqnarray}
Y_e = Y_{e0} + \frac{s}{\lambda}\left(1+\frac{s^2}{\kappa^2}\right).
\end{eqnarray}

The reason for adopting this expression for $Y_e$ is as follows:  we wish to obtain a level crossing at $s = 0$.    We want to be able to dial the derivative of $Y_e$ at $s = 0$ to see how slowly it must vary in order to give adiabatic $\nu\rightleftharpoons\bar{\nu}$ transformation.  This is done by adjusting the parameter $\lambda$:  larger values of $\lambda$ give greater adiabaticity.  Further, if $\lambda$ must be large in order to trigger transformation, it is useful to know if the derivative of $Y_e$ must remain small in order for transformation to continue, or if it can grow at a later time without halting the transformation.  The parameter $\kappa$ controls the scale on which the derivative of $Y_e$ grows away from the location of the level crossing.

In our model, we find that when nonlinear feedback is included, large-scale $\nu\rightleftharpoons\bar{\nu}$ transformation can occur under unexpected conditions.  Provided that the rate of change of $Y_e$ at $s = 0$ is slow enough that some low-energy neutrinos transform, a feedback mechanism begins to operate that tends to keep $H_3$ near zero until a large number of neutrinos have been converted into antineutrinos.

Fig.~(3) illustrates this phenomenon.  As the system approaches resonance, neutrinos begin to convert into antineutrinos.  This causes the neutrino self-interaction potential to decrease.  If the rate of change of the self-interaction potential is large enough, it will overcome the change of the matter potential and push the system back towards resonance.  This feedback is similar to the matter-neutrino resonance described in \cite{Malkus:2014fk}, but in the context of helicity, rather than flavor, transformation.

\begin{figure}
\includegraphics[width=2.8in]{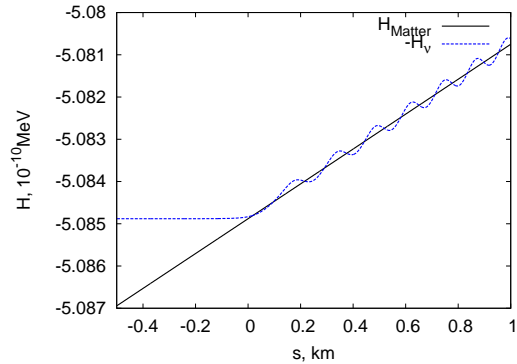}
\caption{Onset of coherent helicity transformation and stabilization of resonance by nonlinear feedback.}
\label{tracking_short}
\end{figure}

If the matter potential changes too quickly, this feedback mechanism fails.  However, with the inclusion of neutrino-neutrino interactions, adiabaticity criteria are much easier to satisfy than in the linear case.  We find that instead of being proportional to $m^{-2}$, as in the linear case, $\lambda$ is proportional to $m^{-4/3}$.  Consequently, inclusion of nonlinear feedback results in the possibility of $\nu\rightleftharpoons\bar{\nu}$ transformation for much faster variation of the matter potential at the level crossing point.

In a supernova environment, $Y_e$ can typically change by $\sim 0.1$ over distances of $\sim 100\,{\rm km}$, so a \lq natural\rq\ value for the scale $\lambda$ is $\sim 1000\,{\rm km}$.  In our model, the required value of $\lambda$ is larger than this even in the presence of nonlinear feedback, except for neutrino masses in excess of $1\,{\rm eV}$.  For example, we find that for $m = 1\,{\rm eV}$, $\lambda \approx 15 \times 1000\,{\rm km}$ is required.  For $m = 0.1\,{\rm eV}$, $\lambda \approx 300 \times 1000\,{\rm km}$ is required.  Cosmological constraints favor neutrino masses not much greater than $0.1\,{\rm eV}$ \cite{Giusarma:2013fk,de-Putter:2012qy,Hannestad:2005uq,Planck-Collaboration:2013fj,Reid:2010kx}, so fine-tuning of the derivative of the matter potential at the level crossing is needed to begin $\nu\rightleftharpoons\bar{\nu}$ transformation.

However, once $\nu\rightleftharpoons\bar{\nu}$ transformation develops, the derivative of the matter potential need not remain unnaturally small.   Fig.~(4) shows the evolution of matter and neutrino Hamiltonians for a model with a slightly exaggerated neutrino mass ($m = 1\,{\rm eV}$), $\lambda = 1.8\times 10^4\,{\rm km}$ and $\kappa = 25\,{\rm km}$.  We see that while a relatively small rate of change of the matter potential is required for transformation to begin, the tracking behavior continues until the rate of change of the matter potential is considerably larger than what it was initially.  This is true even though the helicity-mixing term in the Hamiltonian is proportional to the neutrino lepton number and is decreasing as neutrinos are being converted to antineutrinos.

\begin{figure}
\includegraphics[width=2.8in]{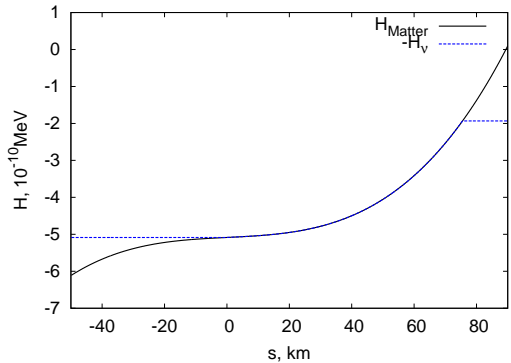}
\caption{Tracking and cancelation of matter potential by neutrino potential over the course of coherent helicity transformation for a model with $m = 1\,{\rm eV}$, $\lambda = 1.8\times 10^4\,{\rm km}$, $\kappa = 25\,{\rm km}$.}
\label{tracking_long}
\end{figure}

The final spectra resulting from helicity transformation in Fig.~(4) are shown in Fig.~(5).  We see that at low energies, neutrinos are converted into antineutrinos, while at higher energies the spectra are relatively unchanged.  This is similar to what is seen in linear MSW.  However, with nonlinear feedback, transformation takes place up to much higher energies than what would be allowed by adiabaticity conditions in linear MSW.

\begin{figure}
\includegraphics[width=2.8in]{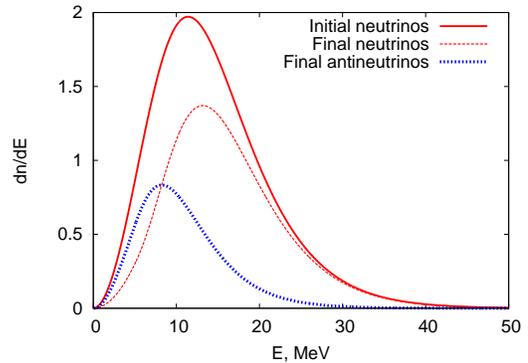}
\caption{Initial and final spectra from a model with $m = 1\,{\rm eV}$, $\lambda = 1.8\times 10^4\,{\rm km}$, $\kappa = 25\,{\rm km}$.}
\label{spectra}
\end{figure}

In conclusion, the QKEs allow the possibility of a level crossing between neutrinos and antineutrinos, potentially resulting in $\nu\rightleftharpoons\bar{\nu}$ transformation.  Conditions required for this level crossing can occur in core collapse supernovae and compact object mergers.

The primary obstacle to helicity transformation is the issue of adiabaticity, which arises due to the fact that the $\nu\rightleftharpoons\bar{\nu}$ mixing term in the Hamiltonian is suppressed by a factor of $m/E\approx 10^{-7}-10^{-8}$. In our toy model, we find that this issue is somewhat alleviated by nonlinear feedback due to neutrino-neutrino interactions, although the derivative of the matter potential at the level crossing must still be smaller than the generically expected value.  In addition, helicity transformation is sensitive to the value of the neutrino mass and becomes much more likely if neutrino masses are larger than their minimum values.

Our model may not fully capture the conditions under which $\nu\rightleftharpoons\bar{\nu}$ transformation can take place.  In a multi-angle system,  there may be additional effects.  The Hamiltonian changes at different rates along different neutrino emission angles, so it is more likely that adiabaticity criteria will be satisfied for some angles, leading neutrinos on these trajectories to transform.  On the other hand, level crossing for different trajectories occurs at different locations, possibly rendering nonlinear feedback ineffective unless a large fraction of neutrinos is nearly collinear.  Also, the $\nu_e\rightleftharpoons\bar{\nu}_e$ resonance can occur when neutrino opacity is not negligible and a significant fraction of neutrinos is propagating inward.  In that case the full multi-angle QKEs cannot be solved by integrating outward in $r$ and must instead be solved as a boundary value or a time evolution problem.  Additionally, a multi-flavor model may exhibit different behavior from the 1-flavor model, as it includes additional resonances and the possibility of exchange of flavor information, in addition to particle number, between neutrinos and antineutrinos.  Finally, we have not considered the effect of the evolution of electron neutrino number on $Y_e$, which could lead to an additional type of feedback \cite{McLaughlin:1999fk,Fetter:2003ys,Wu:2014vn}.

A convincing determination of the extent to which helicity transformation actually takes place in core collapse supernovae or compact object mergers requires sufficiently realistic multi-angle simulations coupled to the evolution of the matter background.  However, given the nonlinear feedback in the QKEs, we cannot preclude significant helicity transformation, with potentially important implications for the physics of compact objects.

This work was supported in part by
NSF grant PHY-1307372 at UCSD, the LDRD Program
at LANL, the University of California Office of the President, the UC HIPACC
collaboration, and by the DOE/LANL Topical Collaboration.

\bibliography{yallref}

\end{document}